\documentclass[12pt]{article}
\usepackage{amsmath,amsthm,amsfonts,amssymb}

\newtheorem{Proposition}{Proposition}
\newtheorem{Definition}{Definition}

\def\Prop #1: #2\par{\medbreak \noindent {\bf Proposititon #1 :}\enspace
{\rm #2}\smallskip\noindent}

\def\Df #1: {\medbreak \noindent {\bf Definition #1} :\enspace}

\def\df #1: #2\par{\smallskip\noindent {\bf #1}\enspace{\bf
Defin\'{\i}ci\'o}:\enspace #2\smallskip\noindent}

\def\Al #1: #2\par{\smallskip\noindent {\bf #1}\enspace
{\bf \'All\'{\i}t\'as}:\enspace #2\smallskip\noindent}

\def\C{$C^{\ast}$-}
\def\W{$W^{\ast}$-}

\def\cA{{\cal A}}
\def\cB{{\cal B}}

\def\cH{{\cal H}}

\def\cK{{\cal K}}

\def\cM{{\cal M}}
\def\cN{{\cal N}}

\def\cS{{\cal S}}

\def\BH{{\cal B} ({\cal H} )}

\def\CC{{\mathbb C}}
\def\IN{{\mathbb N}}
\newcommand{\ie}{{\it i.e.\ }}
\newcommand{\eg}{{\it e.g.\ }}
\newcommand{\etc}{{\it etc}}
\def\As{{\cal A}}
\def\Bs{{\cal B}}

\def\Hs{{\cal H}}

\def\Ms{{\cal M}}
\def\Ns{{\cal N}}
\def\Os{{\cal O}}

\def\Ss{{\cal S}}

\begin{document}

\title{When Are Quantum Systems Operationally Independent?}

\author{{Mikl\'os R\'edei} \\
Department of Philosophy, Logic and Scientific Method  \\
London School of Economics and Political Science\\
Houghton Street, London WC2A 2AE, UK\thanks{e-mail: M.Redei@lse.ac.uk}\\
\vphantom{X}\\
{Stephen J.\ Summers } \\
Department of Mathematics \\ University of Florida \\
Gainesville, FL 32611, USA\thanks{e-mail: sjs@math.ufl.edu}}

\date{April 8, 2009}

\maketitle

{  \abstract \noindent
We propose some formulations of the notion of
``operational independence" of two subsystems $S_1,S_2$ of a larger
quantum system $S$ and clarify their relation to other independence
concepts in the literature. In addition, we indicate why
the operational independence of quantum subsystems holds quite
generally, both in nonrelativistic and relativistic quantum theory. }

\section{Introduction}

     The aim of this note is to propose mathematically well defined
formulations of the notion of ``operational independence" of two
subsystems $S_1,S_2$ of a larger quantum system $S$ and to clarify
their relation to other independence concepts in the mathematical
physics literature.  In addition, we shall indicate why the
operational independence of quantum subsystems holds quite generally,
both in nonrelativistic and relativistic quantum theory.

     Intuitively, operational independence of subsystems $S_1$ and $S_2$
expresses the notion that any two physical operations (measurements, state
preparations \etc) which can be carried out on $S_1$ and $S_2$ {\em
separately\/} can also be carried out {\em jointly\/} as a single
operation on system $S$.

     It will be seen that operational independence can be given
different technical formulations within the context of operator
algebraic models of quantum systems. If the observables of quantum
systems $S_1,S_2$ and $S$ are represented by selfadjoint elements
of \C subalgebras $\cA_1,\cA_2$ of a \C algebra $\cA$, then $S_1$ and
$S_2$ are called {\em operationally \C independent in} $\cA$ if any
two completely positive, unit preserving maps $T_1$ and $T_2$ on
$\cA_1$ and $\cA_2$, respectively, have a joint extension to a
completely positive, unit preserving map $T$ on $\cA$ (Definition
\ref{opinddefc*}). Completely positive maps $T$ satisfying $T(I) \leq I$
are called {\em operations} in the physics literature,
since they can be used to represent
physical operations carried out on the quantum systems \cite{Da,Kr}.
If the observables of the quantum systems in question are represented
by von Neumann algebras, then it is natural to require the operations
$T_1,T_2$ and $T$ to be normal (continuous in the $\sigma$-weak
topology) -- the resulting definition is operational \W independence
(Definition \ref{opinddefw*}). Requiring that the extension T factors
across the subalgebras (and preserves faithfulness) leads to
Definitions \ref{opinddefc*product} and \ref{opinddefw*product}.

     In this paper we shall explain the relations of these notions to
the already established notions of subsystem independence in the
literature and, in so doing, provide some useful alternative
characterizations of operational independence. In addition, we shall
be able to demonstrate that the strongest form of operational
independence formulated here obtains quite generally in
nonrelativistic quantum mechanics and in relativistic quantum field
theory.

     We outline the structure of the paper. Section \ref{Review}
recalls some notions of independence which have been investigated in
the literature and which are relevant from the perspective of
operational independence. Section \ref{cpmaps} recalls the
concept of operation as a completely positive map on \C, resp. \W, algebras
together with some basic properties of completely positive maps. Section
\ref{opinddef} formulates the definitions of operational independence
in terms of completely positive maps and establishes their logical relations
with the notions described in Section \ref{Review}. Finally, in
Section \ref{split} the relation to a further, previously studied
independence property called the split property is explained, and this
relation is used to show that operational independence holds widely in
quantum theory.

\section{Some notions of independence}\label{Review}

     Throughout the paper $\cA$ denotes a unital \C algebra,
$\cA_1,\cA_2$ are assumed to be \C subalgebras of $\cA$ (with common
unit $I$). $\As_1 \vee \As_2$ will denote the smallest \C subalgebra
of $\As$ containing both $\As_1$ and $\As_2$. $\cN$ denotes a von Neumann
algebra, and $\cN_1,\cN_2$ will be von Neumann subalgebras of
$\cN$ (with common unit). $\Ns_1 \vee \Ns_2$ will denote the smallest
von Neumann algebra in $\Ns$ containing both $\Ns_1$ and $\Ns_2$.
If $\Ns$ is a von Neumann algebra acting on the Hilbert space $\Hs$,
then $\Ns{}'$ represents its commutant, the set of all bounded operators
on $\Hs$ which commute with all elements of $\Ns$. $\Ss(\cA)$
is the state space of the \C algebra $\cA$. (For the operator
algebraic notions see \cite{Tak}, \cite{KR} or \cite{Bl}.) For a Hilbert space
$\cH$, the set of all bounded operators on $\cH$ is denoted by $\BH$.

     Since there are different quantitative and qualitative aspects to
the notion of independent subsystems, it is natural that there be many
theory dependent formulations of such independence. We discuss only a
few of these here. The following technical definitions of independence
were formalized in the context of algebraic quantum theory in a
comprehensive review up to 1990 of the hierarchy of independence
concepts and their non-trivial logical interrelations
\cite{Sum90}. See \cite{Sum90} for a discussion of their operational
meaning and their history. For more recent developments, see
\cite{FS,Red,Ham}.

\begin{Definition}\label{defC*independence}
{\rm A pair $(\cA_1,\cA_2)$ of \C subalgebras of a \C algebra $\cA$
is called {\it \C independent} if for any state $\phi_1$ on $\cA_1$
and for any state $\phi_2$ on $\cA_2$ there exists a state $\phi$ on
$\cA$ such that both
\begin{eqnarray*}
\phi(X) & = & \phi_1(X) \mbox{\ \ \ for any\ \ } X \in \cA_1 \\
\phi(Y) & = & \phi_2(Y) \mbox{\ \ \ for any\ \ } Y \in \cA_2
\end{eqnarray*}
obtain.}
\end{Definition}

\begin{Definition}
{\rm A pair $(\cA_1,\cA_2)$ of \C subalgebras of a \C algebra $\cA$
is called {\it \C independent in the product sense} if the map
$\eta(XY) \doteq X \otimes Y$ extends to an \C isomorphism of
$\As_1 \vee \As_2$ with $\As_1 \otimes \As_2$, where $\As_1 \otimes \As_2$
denotes the tensor product of $\As_1$ and $\As_2$ with the minimal
\C norm (see \cite{Tak,KR,EfLa}). 
}
\end{Definition}

     If $\As$ is faithfully represented on a Hilbert space $\Hs$,
then the minimal norm referred to here is the ordinary operator norm in
$\Bs(\Hs) \otimes \Bs(\Hs) \simeq \Bs(\Hs \otimes \Hs)$.

\begin{Definition}\label{defW*independence}
{\rm A pair $(\cN_1,\cN_2)$ of von Neumann subalgebras of the von Neumann
algebra $\cN$ is called {\it \W independent} if for any {\em normal\/}
state\footnote{These are the states which can be represented by a
density matrix. Hence, in general, physicists tacitly restrict their
attention to normal states.}
$\phi_1$ on $\cN_1$ and for any {\em normal\/}  state $\phi_2$ on
$\cN_2$ there exists a {\em normal\/}  state $\phi$ on $\cN$ such that both
\begin{eqnarray*}
\phi(X)&=&\phi_1(X)\mbox{\ \ \ for any\ \ } X\in\cN_1\\
\phi(Y)&=&\phi_2(Y) \mbox{\ \ \ for any\ \ } Y\in\cN_2
\end{eqnarray*}
obtain.}
\end{Definition}

\begin{Definition}
{\rm A pair $(\cN_1,\cN_2)$ of von Neumann subalgebras of the von Neumann
algebra $\cN$ is called {\it \W independent in the product sense} if
for any normal state $\phi_1$ on $\cN_1$ and for any normal state $\phi_2$ on
$\cN_2$ there exists a  normal product state $\phi$ on $\cM$ extending
$\phi_1$ and $\phi_2$, \ie a normal state $\phi$ on $\cN$ such that
$$\phi(XY) = \phi_1(X) \phi_2(Y) \mbox{\ \  for any\ \ \ } X \in \cN_1,
Y \in \cN_2 \, .$$
}
\end{Definition}

     The above independence notions are not independent logically. Here
we collect some results on their interrelations. Note that only
\C independence in the product sense requires that the algebras mutually
commute. The apparent asymmetry between the definitions of \C, resp. \W,
independence in the product sense will be resolved below (for mutually
commuting von Neumann algebras acting on a separable Hilbert space).

\begin{Proposition}\label{summaryprop}
\noindent
\begin{enumerate}

\item If $\cA_1,\cA_2$ are commuting, then the \C independence in the product
sense of $(\cA_1,\cA_2)$ implies the \C independence of $(\cA_1,\cA_2)$, but
the converse is false \cite{Sum90}.

\item \W independence of a pair of arbitrary von Neumann algebras
implies \C independence of the pair \cite{Sum90,FS}, but the converse
is false.  In fact, examples of pairs of von Neumann algebras which do not
mutually commute have been found which are \C independent but not
\W independent.  But if $\cN_1,\cN_2$ are {\em commuting\/} von Neumann
algebras acting on a separable Hilbert space, then the \C independence
of $(\cN_1,\cN_2)$ implies the \W independence of the pair \cite{FS},
so that for such pairs \C independence is equivalent to \W independence.

\item The \W independence in the product sense of $(\Ns_1,\Ns_2)$ implies the
$W^*$-independence of $(\Ns_1,\Ns_2)$, but the converse is false
\cite{Sum90}.

\item If $\cN_1,\cN_2$ are commuting, then the \W independence in the product
sense of $(\Ns_1,\Ns_2)$ implies the \C independence in the product sense
of $(\Ns_1,\Ns_2)$, but the converse is false \cite{Sum90,FS}.
(This is further discussed below.)

\end{enumerate}
\end{Proposition}

     Note that if $\cA_1,\cA_2$ are commuting \C algebras, then the
extension state $\phi$ in Definition \ref{defC*independence} may be
chosen to be a product state \cite{Ro}, \ie
$$\phi(XY) = \phi(X) \phi(Y) = \phi_1(X) \phi_2(Y) \, , $$
for all $X \in \As_1$, $Y \in \As_2$. The corresponding assertion for
\W independence is false \cite{Sum90}. Indeed, in that context
one has the following theorem.

\begin{Proposition}[\cite{Tak58}] \label{TakesakiThm}
Let $\cN_1,\cN_2$ be commuting factor von Neumann algebras acting on a
common Hilbert space $\Hs$. Then the map
$\eta(XY) \doteq X \otimes Y$ extends
to a \W isomorphism of $\Ns_1 \vee \Ns_2$ with the \W tensor product
$\Ns_1 \overline{\otimes} \Ns_2$ if and only if
there exists a normal product state on $\cN_1 \vee \cN_2$.
\end{Proposition}

     In fact, the assumption that the algebras be factors may be dropped
if the normal product state is required to have central support $I$,
the identity map on $\Hs$ \cite{DL}. Hence, one has the following
result.

\begin{Proposition} \label{improved}
Let $\cN_1,\cN_2$ be commuting von Neumann algebras acting on a separable
Hilbert space. Then $(\cN_1,\cN_2)$ is
\W independent in the product sense if and only if there exists a faithful
normal product state on $\Ns_1 \vee \Ns_2$.
\end{Proposition}

\begin{proof}
Let $(\cN_1,\cN_2)$ be \W independent in the product sense. Since the
Hilbert space on which the algebras act is separable, there exist
faithful normal states $\phi_1,\phi_2$ on $\cN_1,\cN_2$, respectively
\cite[Prop. II.3.19]{Tak}. But then $\phi_1 \otimes \phi_2$ is a faithful
normal state on $\Ns_1 \overline{\otimes} \Ns_2$
\cite[Cor. IV.5.12]{Tak}. If
$\eta : \Ns_1 \vee \Ns_2 \to \cN_1 \overline{\otimes} \cN_2$ is the
hypothesized \W isomorphism, then $(\phi_1 \otimes \phi_2) \circ \eta$
is a faithful normal product state on $\Ns_1 \vee \Ns_2$. For the
converse, see \cite{Tak58,DL}.
\end{proof}

     An analogous characterization of \C independence in the product
sense was proven in \cite{FS}.

\begin{Proposition}[\cite{FS}] \label{1997}
Let $\cA_1,\cA_2$ be commuting subalgebras of a \C algebra $\cA$
acting on a separable Hilbert space. Then $(\cA_1,\cA_2)$ is
\C independent in the product sense if and only if there exists a faithful
product state on $\As_1 \vee \As_2$.
\end{Proposition}

     These results resolve the asymmetry between the definitions
of \C, resp. \W, independence in the product sense, at least in the
indicated important special case. It therefore follows that for a pair
of commuting von Neumann algebras acting on a separable Hilbert space,
\W independence in the product sense implies \C independence in the
product sense. However, the converse is false --- see below.

\section{Positive and completely positive maps\label{cpmaps}}

     Recall that a linear map $T \colon \cA \to \cB$ can be extended
to a linear map $T_n \colon M_n(\cA) \to M_n(\cB)$ (here $M_n(\cA)$ is
the set of $n$ by $n$ matrices with entries which are 
elements from the \C algebra $\cA$) by
$$T_n \left(%
\begin{array}{ccc}
  a_{11} & \ldots & a_{1n} \\
  . & . & . \\
  a_{n1} & \ldots & a_{nn} \\
\end{array}%
\right)=
\left(%
\begin{array}{ccc}
  T(a_{11}) & \ldots & T(a_{1n}) \\
  . & . & . \\
  T(a_{n1}) & \ldots & T(a_{nn}) \\
\end{array}%
\right) \, .
$$

\begin{Definition}
{\rm $T$ is {\it completely positive} if $T_n$ is positive for every
$n \in \IN$. A completely positive map $T : \cA \to \As$ satisfying
$T(I) \leq I$ is called an {\em operation} \cite{Da,Kr}. An operation $T$
such that $T(I) = I$ is said to be {\it nonselective}. An operation $T$ on a
von Neumann algebra $\cN$ is called {\em normal\/} if it is
$\sigma$--weakly continuous. A positive  linear map $T : \As \to \Bs$
is {\it faithful} if $T(X) > 0$ whenever $\As \ni X > 0$.
}
\end{Definition}

     The dual $T^*$ of a nonselective operation defined by
$$T^* \colon \Ss(\cA) \to \Ss(\cA) \qquad\qquad T^* \phi \doteq \phi \circ T $$
maps the state space $\Ss(\cA)$ of $\cA$ into itself. If $T$ is a normal
nonselective operation on the von Neumann algebra $\cN$, then $T^*$
takes normal states to normal states.

     Operations are the mathematical representatives of physical
operations, \ie physical processes which take place as a result of
physical interactions with the quantum system. (For a detailed
interpretation of operations see \cite{Kr}.) A state on $\As$ is a completely
positive unit preserving map from $\As$ to $\CC$ \cite{Arv}. So, if
$\phi$ is a state on $\As$, then
\begin{equation}
\As \ni X\mapsto T(X) = \phi(X) I \in \As
\end{equation}
is a nonselective operation in the sense of the above
definition, which is canonically associated with the state and which
may be interpreted as the preparation of the system into the state
$\phi$. Further examples of operations are provided by
measurements. In particular, if one measures a quantum system with
observable algebra
$\Bs(\Hs)$ for the value of a (possibly unbounded) observable $Q$ with
purely discrete spectrum $\{ \lambda_i \}$ and corresponding spectral
projections $P_i$, then according to the ``projection postulate" this
measurement can be represented by the operation $T$ defined as
\begin{equation} \label{measure}
\Bs(\Hs) \ni X \mapsto T(X) = \sum_iP_iXP_i \in \Bs(\Hs) \, .
\end{equation}
$T$ is a normal nonselective operation.

     A classic result characterizing certain completely positive maps was
established in \cite{St}.

\begin{Proposition} [Stinespring's Representation Theorem]
{\rm $T \colon \cA \to \BH$ is a completely positive linear map from a
\C algebra $\cA$ into $\BH$ if and only if it has the form
$$T(X) = V^* \pi(X) V \qquad\qquad X \in \cA \, , $$
where $\pi \colon \cA \to \cB(\cK)$ is a representation of $\cA$ on the
Hilbert space $\cK$ and $V \colon \cH \to \cK$ is a bounded linear map. If
$\cA$ is a von Neumann algebra and $T$ is normal, then $\pi$ can be
chosen to be a normal representation.
}
\end{Proposition}

     So, in particular, \C homomorphisms are completely positive. A
corollary of Stinespring's theorem was proven by Kraus \cite{Kr}.

\begin{Proposition} [Kraus' Representation Theorem]
{\rm $T \colon \BH \to \BH$ is a normal operation if and only if there
exist bounded operators $W_i$ on $\cH$ such that
$$T(X) = \sum_i W_i^* X W_i  \qquad\qquad \sum_i W_i^* W_i \leq I \, . $$
}
\end{Proposition}

\noindent Compare with equation (\ref{measure}).

     It is important in Stinespring's theorem that $T$ takes its value
in the set of all bounded operators $\BH$ on a Hilbert space.
This is related
to the fact that operations defined on a subalgebra of an arbitrary \C
algebra are {\em not\/}, in general, extendible to an operation
on the larger algebra \cite{Arv}. Indeed, a \C algebra $\Bs$ is said
to be {\it injective} if for any \C algebras $\As_1 \subset \As$
every completely positive unit preserving linear map
$T_1 : \As_1 \to \Bs$ has an extension to a completely positive unit
preserving linear map $T : \As \to \Bs$. It was shown in \cite{Arv}
that $\Bs(\Hs)$ is injective.

\section{Operational independence\label{opinddef}}

     In the light of these considerations, the following generalizations
of \C and \W independence are natural.

\begin{Definition}\label{opinddefc*}
{\rm A pair $(\cA_1,\cA_2)$ of \C subalgebras of \C algebra $\cA$ is
{\it operationally \C independent in $\cA$} if any two nonselective operations
on $\cA_1$ and $\cA_2$, respectively, have a joint extension to a
nonselective operation on $\cA$; \ie if for any two completely
positive unit preserving maps
$$T_1 \colon \cA_1 \to \cA_1 \quad , \quad
T_2 \colon \cA_2 \to \cA_2 \, , $$
there exists a completely positive unit preserving map
$$T \colon \cA \to \cA$$
such that
\begin{eqnarray*}
T(X) & = & T_1(X) \qquad \mbox{for all\ } X \in \cA_1 \\
T(Y) & = & T_2(Y) \qquad \mbox{for all\ } Y \in \cA_2 \, .
\end{eqnarray*}
}
\end{Definition}

\begin{Definition}\label{opinddefw*}
{\rm A pair $(\cN_1,\cN_2)$ of von Neumann subalgebras of a von Neumann
algebra $\cN$ is {\it operationally \W independent in $\cN$} if any two
{\em normal\/} nonselective operations on $\cN_1$ and $\cN_2$, respectively,
have a joint extension to a {\em normal\/} nonselective operation on $\cN$.
}
\end{Definition}

\noindent Since operations defined on a subalgebra need not be extendible to
a larger algebra in general, it is important in Definitions
\ref{opinddefc*} and \ref{opinddefw*} that operational independence
of subalgebras is defined with respect to some fixed larger algebra.
Note, however, that, here and below, this joint extension then has
further extensions to arbitrary superalgebras, as long as the range of the
first extension is interpreted as mapping into an injective algebra,
which remains the fixed range of the further extensions.

     Operational \C independence expresses the notion that any
operation (measurement, state preparation \etc) on system
$S_1$ is co-possible with any such operation on system $S_2$ (if these
systems are represented by \C algebras --- similarly for \W
algebras). Given a nonselective operation $T$, its dual $T^*$
takes states into states; hence, the content of operational \C and \W
independence also can be formulated in terms of changes of states of
the systems involved: Operational \C independence of $(\cA_1,\cA_2)$
entails the feature that any {\em transition\/} of state
$\phi_1$ of $\cS_1$ into state $\psi_1$ is compatible with any
{\em transition\/} $\phi_2$ of $\cS_2$ into state $\psi_2$. That is to say,
these two transitions can take place as a transition of a single state
$\phi$ of $\cS$ into state $\psi$.  Operational \W independence has a
similar interpretation in terms of transitions between {\em normal\/}
states on the respective von Neumann algebras.

    In analogy with \C and \W independence in the product sense, the
following strengthened versions of operational \C and \W independence
seem natural.

\begin{Definition}\label{opinddefc*product}
{\rm A pair $(\cA_1,\cA_2)$ of \C subalgebras of a \C algebra $\cA$ is
{\it operationally \C independent in $\cA$ in the product sense}
if any two (faithful) nonselective operations on $\cA_1$ and $\cA_2$,
respectively, have a joint extension to a (faithful) nonselective
operation on $\cA$ which is a {\em product} across $\cA_1$ and $\cA_2$;
\ie if for any two (faithful) completely positive unit preserving maps
$$T_1 \colon \cA_1 \to \cA_1 \quad , \quad
T_2 \colon \cA_2 \to \cA_2 \, ,$$
there exists a (faithful) completely positive unit preserving map
$$T\colon\cA\to\cA$$
such that
\begin{eqnarray}
T(X) & = & T_1(X) \qquad \mbox{for all\ } X \in \cA_1 \\
T(Y) & = & T_2(Y) \qquad \mbox{for all\ } Y \in \cA_2 \\
T(XY)& = & T(X)T(Y) \qquad X \in \cA_1 \qquad Y \in \cA_2 \label{product}
\end{eqnarray}
}
\end{Definition}

\begin{Definition}\label{opinddefw*product}
{\rm A pair $(\cN_1,\cN_2)$ of von Neumann subalgebras of a von Neumann
algebra $\cN$ is {\it operationally \W independent in $\cN$ in the product
sense} if any two (faithful) {\em normal\/} nonselective
operations on $\cN_1$ and $\cN_2$, respectively, have a joint extension to
a (faithful) {\em normal\/} nonselective
operation $T$ on $\cN$ which is a {\em product} across $\cN_1$ and $\cN_2$
in the sense of eq. (\ref{product}).
}
\end{Definition}

     We first remark that in Definition \ref{opinddefw*product} the
{\it prima facie} additional requirement that faithful operations are
extended by faithful operations is superfluous in the case of states.
In other words, \W independence in the product sense {\it entails}
that faithful states can be extended by faithful product states
(cf. the proof of Prop. \ref{improved}). This is not true in the case
of states in Definition \ref{opinddefc*product} \cite{Hampriv}. The status of
this additional requirement is under investigation in the case of
general operations \cite{Hampriv}. The assumption is added here for
reasons which will become apparent below.

     States provide special cases of operations, yet \C and
\W independence are {\em not\/}, strictly speaking, special cases of
operational \C and \W independence. Indeed, \C and \W independence
require a narrower class of operations on $\cS_1$ and $\cS_2$ to have a
joint extension, but the joint extension must belong, in turn, to that
narrower class of operations (the states). On the other hand, operational
\C and \W independence require a larger class of partial operations to have
a joint extension, but the extension can be in that larger class of operations.
Thus \C and \W independence on one hand, and operational \C and \W
independence on the other, are {\it prima facie} not related in a
straightforward manner. Let us examine this relationship more closely.
Assume that $(\cA_1,\cA_2)$ is operationally \C independent in $\cA$.
Let $\phi_1$ and $\phi_2$ be two states on $\cA_1$ and $\cA_2$,
respectively. As mentioned earlier, the two maps
\begin{eqnarray}
T_1(X) & = & \phi_1(X) I \,  , \, X \in \As_1 \, , \label{1} \\
T_2(Y) & = & \phi_2(Y) I \,  , \, Y \in \As_2 \, , \label{2}
\end{eqnarray}
are completely positive unit preserving maps on $\cA_1$ and $\cA_2$,
respectively, so by assumption, $T_1$ and $T_2$ have a
joint extension $T$ to $\cA$. This $T$ need not be associated with a
state; however, for {\em any} state $\phi$ on $\cA$, the state
$T^*\phi$ on $\As$ is clearly an extension of both $\phi_1$ and $\phi_2$.
It is clear that similar reasoning remains valid if the states
$\phi_1,\phi_2$ and $\phi$ are assumed to be normal states on operationally
\W independent von Neumann subalgebras $\cN_1$ and $\cN_2$ of $\cN$. What
is more, if operational independence in the product sense obtains, then
one has
\begin{equation} \label{productstate}
T^* \phi(XY) = \phi(T(XY)) = \phi(T(X)T(Y)) = \phi(T_1(X) T_2(Y)) =
\phi_1(X) \phi_2(Y) \, ,
\end{equation}
for all $X \in \As_1$, $Y \in \As_2$, and for {\it any} state
$\phi \in \Ss(\As)$.  We observe that operational independence in the product
sense thereby entails the existence of operations which prepare the
quantum system presented in any initial (normal) state into a product
state yielding any two prescribed (normal) partial states. This is a
remarkable property; therefore it is noteworthy that operational
independence in the product sense can be verified in rather general
circumstances (see the next section). In light of these remarks, we
have a series of propositions; the proofs of the first two are now
immediate.

\begin{Proposition}\label{entailmentsc*}
Operational \C independence of $(\cA_1,\cA_2)$ in $\cA$ entails the
\C independence of the pair $(\cA_1,\cA_2)$.
\end{Proposition}

\begin{Proposition}\label{entailmentsw*}
Operational \W independence of $\cN_1$ and $\cN_2$ in $\cN$ entails the
\W independence of the pair $(\cN_1,\cN_2)$.
\end{Proposition}

\noindent Note that in Propositions \ref{entailmentsc*} and
\ref{entailmentsw*} the algebras $(\cA_1,\cA_2)$ and $(\cN_1,\cN_2)$ are
{\em not\/} assumed to be commuting.

     Before proceeding to the next results, we need the following
proposition. A proof of most, but not all, of the assertions in this
proposition using the Stinespring representation theorem can be found in
\cite[Lemma 2.5]{EfLa}. We present an alternative argument here which
also establishes the remaining points.

\begin{Proposition} [\cite{EfLa}] \label{faithful}
Let $\As_1,\As_2,\Bs_1,\Bs_2$ be unital \C algebras and let
$T : \As_1 \to \Bs_1$ and \newline
$S : \As_2 \to \Bs_2$ be (faithful) completely positive maps. Then
$T \otimes S : \As_1 \otimes \As_2 \to \Bs_1 \otimes \Bs_2$ is a (faithful)
completely positive map. If $\As_1,\As_2,\Bs_1,\Bs_2$ are von Neumann
algebras and $T$ and $S$ are normal, then
$T \otimes S : \As_1 \overline{\otimes} \As_2 \to
\Bs_1 \overline{\otimes} \Bs_2$ is normal.
\end{Proposition}

\begin{proof}
That $T \otimes S$ is completely positive, resp. normal, under the
stated conditions is a consequence of \cite[Prop. IV.4.23, Prop. IV.5.13]{Tak}.
So let $S$ and $T$ be faithful and $0 \neq A \in \As_1 \otimes \As_2$.
Let $\hat{I}_{\As_1}$, resp. $\hat{I}_{\As_2}$ \etc, denote the identity
map on $\As_1$, resp. $\As_2$ \etc. These maps are completely positive.

     First, consider the case $\As_2 = \Bs_2$. By \cite[Thm. IV.4.9]{Tak}
there exist $\phi_1 \in \Ss(\As_1)$, $\phi_2 \in \Ss(\As_2)$, such that
$(\phi_1 \otimes \phi_2)(AA^*) \neq 0$. Let $T_1,T_2$ be the completely
positive maps defined in (\ref{1}), (\ref{2}). Since
$\hat{I}_{\As_1} \otimes T_2$ is completely positive and $AA^*$ is positive,
one must have $(\hat{I}_{\As_1} \otimes T_2)(AA^*) \geq 0$. And since
$$(\phi_1 \otimes \phi_2)(\hat{I}_{\As_1} \otimes T_2)(AA^*) =
(\phi_1 \otimes \phi_2)(AA^*) \neq 0 \, , $$
one must also have $(\hat{I}_{\As_1} \otimes T_2)(AA^*) \neq 0$.
One therefore concludes $(\hat{I}_{\As_1} \otimes T_2)(AA^*) > 0$.
Note that $(\hat{I}_{\As_1} \otimes T_2)(AA^*)$ can be naturally identified
with a strictly positive element of $\As_1$ as follows. Given the state
$\phi_2$ on $\As_2$, one has the left slice map
$L : \As_1 \otimes \As_2 \to \As_1$ which satisfies
$$L(\sum_i X_i \otimes Y_i) = \sum_i \phi_2(Y_i)X_i \, .$$
This map is completely positive \cite[II.9.7.1]{Bl}, and one has
$(\hat{I}_{\As_1} \otimes T_2)(AA^*) = L(AA^*) \otimes I_{\As_2}$,
where $I_{\As_2}$ is the unit in $\As_2$. Therefore,
$L(AA^*) > 0$. But then
$$(\hat{I}_{\Bs_1} \otimes T_2) \circ (T \otimes \hat{I}_{\As_2})(AA^*) =
(T \circ L(AA^*)) \otimes I_{\As_2} > 0 \, , $$
since $T$ is faithful. This entails that
$(T \otimes \hat{I}_{\As_2})(AA^*) \neq 0$ and thus
$T \otimes \hat{I}_{\Bs_2}$ is faithful (recall $I_{\As_2} = I_{\Bs_2}$ here).
A similar argument implies that $\hat{I}_{\As_1} \otimes S$ is faithful in
the case $\As_1 = \Bs_1$.

     In the general case, one notes that
$T \otimes S = (T \otimes \hat{I}_{\Bs_2}) \circ (\hat{I}_{\As_1} \otimes S)$,
and the proposition follows.
\end{proof}

     An immediate consequence of this observation is given next.

\begin{Proposition}\label{goodc}
Let $\As_1,\As_2$ be mutually commuting \C algebras acting on a separable
Hilbert space. The pair $(\cA_1,\cA_2)$ is \C independent in the product
sense if and only if it is operationally \C independent in
$\As_1 \vee \As_2$ in the product sense.
\end{Proposition}

\begin{proof}
     Let $(\cA_1,\cA_2)$ be \C independent in the product sense, so there
exists a \C isomorphism $\eta : \As_1 \vee \As_2 \to \As_1 \otimes \As_2$
such that $\eta(XY) = X \otimes Y$, for all $X \in \As_1$ and $Y \in \As_2$.
If $T_i$ is a (faithful) completely positive unit preserving map on
$\As_i$, $i = 1,2$, then $T_1 \otimes T_2$ is a (faithful) completely
positive unit preserving map on $\As_1 \otimes \As_2$. Thus,
$(T_1 \otimes T_2) \circ \eta$ is such a map on $\As_1 \vee \As_2$ and
satisfies all the conditions required to establish the operational \C
independence in $\As_1 \vee \As_2$ in the product sense of $(\cA_1,\cA_2)$.

     Conversely, let $(\cA_1,\cA_2)$ be operationally \C independent in
$\As_1 \vee \As_2$ in the product sense. There exist faithful states
$\phi_1,\phi_2$ on $\As_1,\As_2$, respectively (there exist such states
on $\As_1{}''$ and $\As_2{}''$ by \cite[Prop. II.3.19]{Tak} --- just
restrict these to $\As_1$ and $\As_2$, respectively), so that
$T_1,T_2$ defined as in (\ref{1}) and (\ref{2}) are faithful operations
on $\As_1,\As_2$, respectively. By hypothesis, there exists a faithful
joint product extension $T$ on $\As_1 \vee \As_2$. Choosing the
state $\phi$ in equation (\ref{productstate}) to be faithful on
$\As_1 \vee \As_2$, one then has a faithful product state on
$\As_1 \vee \As_2$. Prop. \ref{1997} completes the proof.
\end{proof}

     Of course, a similar argument yields the analogous result in the
\W case.

\begin{Proposition}\label{goodw}
Let $\Ns_1,\Ns_2$ be mutually commuting von Neumann algebras acting on
a separable Hilbert space. The pair $(\cN_1,\cN_2)$ is \W independent in
the product sense if and only if it is operationally \W independent
in $\Ns_1 \vee \Ns_2$ in the product sense.
\end{Proposition}

     In light of Propositions \ref{summaryprop}, \ref{goodc} and
\ref{goodw}, we can then conclude that operational \W independence
in the product sense is strictly stronger than operational
\C independence in the product sense. In fact, choosing $\Ns_1$ to be the
hyperfinite type III factor\footnote{See \cite{KR,Tak} for a
description of the Murray--von Neumann classification of von Neumann
algebras and subsequent refinements. See also \cite{RS} for a discussion
of the necessity and physically relevant consequences of the various
types of von Neumann algebras in quantum theory.}
and $\Ns_2 = \Ns_1{}'$, the pair $(\Ns_1,\Ns_2)$ is \C independent in
the product sense, but it is not \W independent in the product sense
\cite{Sum90,FS}. (This situation actually arises in relativistic quantum
field theory --- cf. \eg \cite{Sum90}.)

\begin{Proposition}
Let $\Ns_1,\Ns_2$ be mutually commuting von Neumann algebras acting on
a separable Hilbert space. For the pair $(\Ns_1,\Ns_2)$, operational
\W independence in $\Ns_1 \vee \Ns_2$ in the product sense implies
operational \C independence in $\Ns_1 \vee \Ns_2$ in the product
sense, but the converse is false.
\end{Proposition}

\section{Operational independence and the split property} \label{split}

     In this section we discuss the relation of operational independence
with a further well studied independence property and use this relation
to demonstrate that operational \W independence in the product sense
holds quite generally in both nonrelativistic and relativistic quantum
theory. The independence property in question is a strengthening of
\W independence in the product sense.

\begin{Definition}
{\rm A pair $(\cN_1,\cN_2)$ of von Neumann subalgebras acting on a Hilbert
space $\Hs$ is called {\it \W independent in the spatial product sense} if
the map
$$XY \to X \otimes Y \qquad X \in \cN_1 \quad Y \in \cN_2 $$
extends to a spatial isomorphism of $\cN_1 \vee \cN_2$ with
$\cN_1 \overline{\otimes} \cN_2$, \ie there exists a unitary
operator $U : \Hs \to \Hs \otimes \Hs$ such that
$U XY U^* = X \otimes Y$ for all $X \in \cN_1$, $Y \in \cN_2$.
}
\end{Definition}

     In general, \W independence in the spatial product sense is
strictly stronger than \W independence in the product sense
\cite{DL}. However, there are many commonly met situations in which
they are equivalent \cite[Thm. 1, Cor. 1]{DL}, in particular when
either of the von Neumann algebras is a factor or either is of type
III. \W independence in the spatial product sense is, in turn, known
to be equivalent to an
important structure property of inclusions of von Neumann algebras,
which has been intensively studied for the purposes of both abstract
operator algebra theory and algebraic quantum field theory.

\begin{Proposition} [\cite{Bu}] For a mutually commuting pair
$(\cN_1,\cN_2)$ of von Neumann algebras, the following are equivalent.

   1. There exists a type I factor $\Ms$ such that
$\Ns_1 \subset \Ms \subset \Ns_2{}'$.

   2. $(\cN_1,\cN_2)$ is \W independent in the spatial product sense.

\end{Proposition}

\noindent Although according to the usage introduced in \cite{DoL} we
should say that the pair $(\Ns_1,\Ns_2{}')$ is split, it is for our
purposes more convenient to say that a pair $(\cN_1,\cN_2)$ of von
Neumann algebras is {\it split} if condition (1) in the previous
proposition holds.

     As a consequence of the results discussed above, it is now
evident that operational \W independence in the product sense obtains
in many physically relevant settings. In order not to lengthen this
note unduly, we shall make some brief comments and not formulate
specific theorems.\footnote{However, some of the matters discussed
in this section are treated in more detail in \cite{Sum08}.} 

     In nonrelativistic quantum mechanics, the algebras of observables
are typically type I factors; therefore in that setting mutually
commuting algebras of observables are necessarily split. Hence,
such pairs of algebras are operationally \W independent in the product
sense.

     In relativistic quantum theory \cite{Ara,Haa}, where the algebra
of observables $\As(\Os)$ carries the interpretation of the algebra
generated by all observables measurable in the spacetime region $\Os$,
the local algebras $\As(\Os)$ are typically type III von Neumann
algebras \cite{Fr,BDF}. Hence, for spacelike separated spacetime
regions $\Os_1,\Os_2$ (for which $\As(\Os_1)$ and $\As(\Os_2)$ mutually
commute), the operational \W independence in the product
sense of $(\As(\Os_1),\As(\Os_2))$ is {\it equivalent} to the pair
being split. In \cite{BDL,We} it has been shown that, in the presence
of the additional structures present in algebraic quantum field
theory, the split property is equivalent to the local preparability of
arbitrary normal states on the local algebras; this latter involves a
special case of the operation (\ref{1}) (cf. also
\cite[Thm. 3.13]{Sum90} for a formulation which does not require those
additional structures). Hence, the equivalences we have established
above are not unexpected.

     The split property has been verified for all {\it strictly}
spacelike separated\footnote{The regions remain spacelike separated
even under translation by a sufficiently small neighborhood of the
origin.}  (precompact, convex) regions $\Os_1,\Os_2$ in a number of
physically relevant quantum field models, both interacting and
noninteracting \cite{Bu,Sum82}.\footnote{It is known to fail in some
physically pathological models, such as models with noncompact global
gauge group and models of free particles such that the number of
species of particles grows rapidly with mass \cite{DoL}.} Moreover, the
split property for all strictly spacelike separated (precompact,
convex) regions $\Os_1,\Os_2$ has also been shown to be a consequence
of a condition (nuclearity) which expresses the requirement that the
energy--level density for any states essentially localized in a bounded
spacetime region cannot grow too fast with the energy and assures
that the given model is thermodynamically well--behaved (\eg thermal
equilibrium states exist for all temperatures \cite{BW,BJ}). Hence,
for such regions the pair $(\As(\Os_1),\As(\Os_2))$ of observable
algebras typically satisfies operational \W independence in the
product sense. On the other hand, in general, pairs
$(\As(\Os_1),\As(\Os_2))$ associated with regions which are spacelike
separated and tangent are not \W independent in the product sense
\cite{SW,Sum90} (although they are \W independent) and therefore not
operationally \W independent in the product sense.  Moreover, pairs
$(\As(\Os_1),\As(\Os_2))$ associated with certain unbounded spacelike
separated regions (\eg wedges) cannot be split \cite{Bu} and thus are not
operationally \W independent in the product sense.

\bigskip
\noindent
{\bf Acknowledgement}: Work supported in part by the Hungarian Scientific Research Found (OTKA), contract number: K68043.

\end{document}